\documentclass[useAMS,usenatbib]{mn2e}
\usepackage{graphicx}     
\usepackage{epsfig}
\usepackage{times}    
\usepackage{subfigure}

\title[Cosmological parameters from lens surveys]{Estimating cosmological parameters from future gravitational lens surveys  }
\author[Dobke et al.]{B. M. Dobke$^{1,2,3} $\thanks{E-mail:
benjamin.m.dobke@jpl.nasa.gov; bmd@astro.caltech.edu},  L. J. King$^{2}$\thanks{E-mail: ljk@ast.cam.ac.uk},  C. D. Fassnacht$^{4}$ and M. W. Auger$^{4,5}$\\$^{1}$NASA Jet Propulsion Laboratory, 4800 Oak Grove Drive, Pasadena, CA 91109, USA \\$^{2}$Institute of Astronomy, University of Cambridge, Madingley Road, Cambridge, CB3 0HA, UK \\$^{3}$Dark Cosmology Centre, Niels Bohr Institute, University of Copenhagen, Juliane Maries Vej 30, 2100, Copenhagen, Denmark \\$^{4}$Department of Physics, University of California, 1 Shields Avenue, Davis, CA 95616, USA \\$^{5}$Department of Physics, University of California, Santa Barbara, CA 93106, USA}
\date{Accepted 2009 April 3. Received 2009 April 2; in original form 2007 November 29}
\pagerange{\pageref{firstpage}--\pageref{lastpage}} \pubyear{2009}

\begin{document}

\label{firstpage}

\maketitle

\begin{abstract}
Upcoming ground- and space-based observatories such as the DES, the LSST, the JDEM concepts and the SKA, promise to dramatically increase the size of strong gravitational lens samples.  
A significant fraction of the systems are expected to be time delay lenses. Due to the sensitivity of time delays to cosmological parameters, this increase in sample size opens up new avenues of parameter constraint. Many of the existing lensing degeneracies, e.g. between the Hubble parameter $H_{0}$ and the slope of the density profile of the lens, become less of an issue with large samples since the distributions of a number of parameters are predictable, and can be incorporated into an analysis, thus helping to lessen the degeneracy.  Assuming a mean galaxy density profile that does not evolve with redshift, a $\Lambda$CDM cosmology, and Gaussian distributions for bulk parameters describing the lens and source populations, we generate synthetic lens catalogues and examine the relationship between constraints on the $\Omega_{\rm m}$ -- $\Omega_{\Lambda}$ plane and $H_{0}$ with increasing lens sample size.  We find that, with sample sizes of $\sim$ 400 time delay lenses, useful constraints can be obtained for $\Omega_{\rm m}$ and $\Omega_{\Lambda}$ with approximately similar levels of precision as from the best of other methods.  In addition, sample sizes of $\sim$100 time delay systems yield estimates of $H_{0}$ with errors of only a couple of percent, exceeding the level of precision from current best estimates such as the HST Key Project.  As such, strong lensing promises to become one of the premier methods for constraining the Hubble constant in the future.  We note that insufficient prior knowledge of the lens samples employed in the analysis, via under or overestimates in the mean values of the sample distributions, results in broadening of constraints. This highlights the need for sound prior knowledge of the sample before useful cosmological constraints can be obtained from large time delay samples.

\end{abstract}


\section{Introduction}

Strong gravitational lensing provides us with a unique and compelling probe of cosmology.  In particular, it has been known for some years that time delays between multiple images of certain strong lensing systems allows for constraints on $H_{0}$, the Hubble constant, \citep{refs}, due to the dependence of the time delay scaling on this parameter.  The constraints on $H_{0}$ by means of strong lensing time delays has historically had mixed success, mainly due to its degeneracy with the matter density profile of the lensing galaxy (e.g. \citealt{falc}; \citealt{gore}; \citealt{corl}).  However, despite a number of remaining systematic uncertainties, we have recently witnessed increasing precision from this method (e.g. \citealt{saha}; \citealt{ogur}) together with general agreement with the HST Key Project value of 
$72 \pm 8\;{\rm km}\,{\rm s}^{-1}\,$Mpc$^{-1}$ (\citealt{free}).

Lens systems and their time delays, however, are also sensitive to other cosmological parameters, specifically the matter density of the Universe, $\Omega_{\rm m}$, the dark energy density $\Omega_{\rm DE}$, the dark energy equation of state parameter, \emph{w}, and the curvature $\Omega_{K}$.  Our current best understanding of the universe is that \emph{w}  = -1 (e.g. \citealt{sper}), consistent with dark energy as a cosmological constant; from hereon we refer to $\Omega_{\rm DE}$ as $\Omega_{\Lambda}$.  Together with $H_{0}$, these parameters enter lens models through the angular diameter distances between the source, lens, and observer, as discussed below.

Previous analyses of lens statistics have already yielded useful constraints on $\Omega_{\Lambda}$.  A combination of the Cosmic Lens All-Sky Survey (CLASS; \citealt{brow}; \citealt{myer}) with recent Sloan Digital Sky Survey (SDSS) data on the local velocity dispersion distribution function of E/S0 galaxies (\citealt{bern2}) was used to provide a constraint of $\Omega_{\Lambda}$ $\sim 0.74 - 0.78$ (assuming flatness), and a 95\% confidence upper bound of $\Omega_{\Lambda}$ $<$ 0.86, in good agreement with results from the redshift-magnitude relation of Type Ia supernovae and from CMB experiments \citep{mitc}. 

 The sensitivity of the time delays is at its strongest when considering the Hubble constant, allowing useful constraints to be obtained even when analysing only one system (see e.g. \citealt{cour}; \citealt{kochb}; \citealt{jack} for a review of $H_{0}$ estimates from lensing).  This is not the case for the remaining cosmological parameters however, since the sensitivity of the time delays to $\Omega_{\rm m}$, $\Omega_{\Lambda}$, and $\Omega_{K}$ is relatively weak.  This results in the normal practice of fixing these parameters to the concordance cosmology as derived from a combination of measurements (e.g. SN1a \citealt{tonr}; CMB \citealt{sper}), and focusing on using lenses to constrain $H_{0}$.

The situation could quickly change in the near future. The past two decades have witnessed a golden age of astronomy and cosmology, with observatories such as the Hubble Space Telescope (HST), Chandra, Keck, the Very Large Telescope (VLT), Very Large Array (VLA), and the Wilkinson Microwave Anisotropy Probe (WMAP), to name but a few, providing us with discoveries that have changed our perception of the cosmos from the smallest to the largest scales.  However, we are now beginning to enter a new era, as these great observatories and instruments are either replaced or augmented by a next generation of space- and ground-based telescopes.  Included amongst these are the Large Synoptic Survey Telescope (LSST; \citealt{ivez}), GAIA \citep{perr}, Dark UNiverse Explorer (DUNE; \citealt{refr}), the European Extremely Large Telescope (E-ELT; \citealt{gilm}), the Thirty Meter Telescope (TMT; \citealt{silv}), the Joint Dark Energy Mission (e.g. DESTINY (\citealt{benf}), ADEPT \footnote{See http://universe.nasa.gov/program/probes/adept.html}, and SNAP (\citealt{jeli}) concepts), and the Square Kilometre Array (SKA; \citealt{cari}).  This range of instruments, and others not mentioned above, will have a direct impact on the cosmology that can be done with strong lensing. Predictions for the number of galaxy-scale strong lensing candidates vary between the instruments, but conservative estimates put the combined tally into the many tens of thousands.  One particular example, the proposed NASA/DOE Joint Dark Energy Mission concept SNAP, is predicted to discover between 5,000-50,000 lensed galaxies, and 100-1,000 lensed quasars \citep{mars}, a significant fraction of the latter having corresponding time delays (currently $\sim 15\%$ of known lens systems are time delay lenses).  Some telescopes will make discoveries of lens systems fortuitously, in much the same way as is being done with analysis of the Sloan Digital Sky Survey (SDSS) data, while others will have dedicated programs searching for lens candidates -- notably the LSST with its high cadence observations providing time delay monitoring on a sub-weekly basis.  Instruments such as the Square Kilometre Array, which lie a little further in the future, could also yield thousands of lens systems \citep{koop2}.  The proposed $30-100$m-class ELTs and the TMT will enable the lens galaxies themselves to be studied in exquisite detail, helping to constrain their velocity dispersions and density distributions.  We highlight that samples provided by the above instruments, even when considered purely on an individual basis, will dwarf the current samples of $\sim$ 20 time delay lenses.

This prompts the question as to how such an increased sample of time delays will be used to 
benefit cosmological studies, in particular in the constraint of cosmological parameters such as $\Omega_{\rm m}$, $\Omega_{\Lambda}$ and $H_{0}$.  Indeed, while it is true that the angular diameter distances are only weakly sensitive to the cosmological parameters $\Omega_{\rm m}$, $\Omega_{\Lambda}$, and $\Omega_{K}$, a vastly increased sample size would compensate for this factor.  Additionally, many of the existing lensing degeneracies, e.g. between $H_{0}$ and the slope of the density profile of the lens, should become less of an issue with large samples of lenses since many parameters should tend towards a distribution that can be incorporated into an analysis. 
This would provide another independent constraint on those cosmological parameters, and in combination with other methods, using CMB measurements, SN1a, and X-ray clusters, would result in further increases in precision.  Hence, the aim of this work is to investigate the relationship between constraints on the $\Omega_{\rm m}$ -- $\Omega_{\Lambda}$ plane and on $H_{0}$ as a function of increasing lens time delay sample size.

This paper is organised as follows.  We begin in \S2 by describing the generation of large synthetic data sets and justifying the exact distributions of parameters used.  \S3 will describe how we then take these data sets and analyse them to obtain meaningful constraints on the cosmological parameters discussed.  \S4 presents the key findings from our investigation, highlighting constraints in the $\Omega_{\rm m}$ -- $\Omega_{\Lambda}$ plane and briefly examining the impact that increased sample size can also have on the determination of $H_{0}$.  Finally, \S5 will provide a discussion of our results and present conclusions.

\section{A Synthetic Lens Sample}

\begin{figure*}
\epsfig{file= 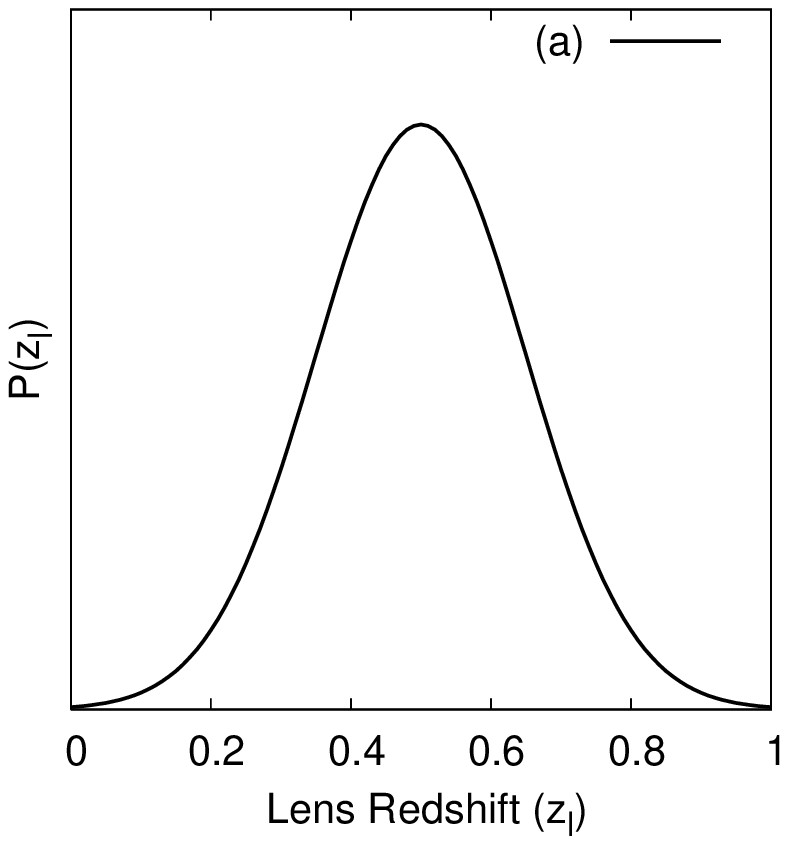,scale=0.448}
\epsfig{file= 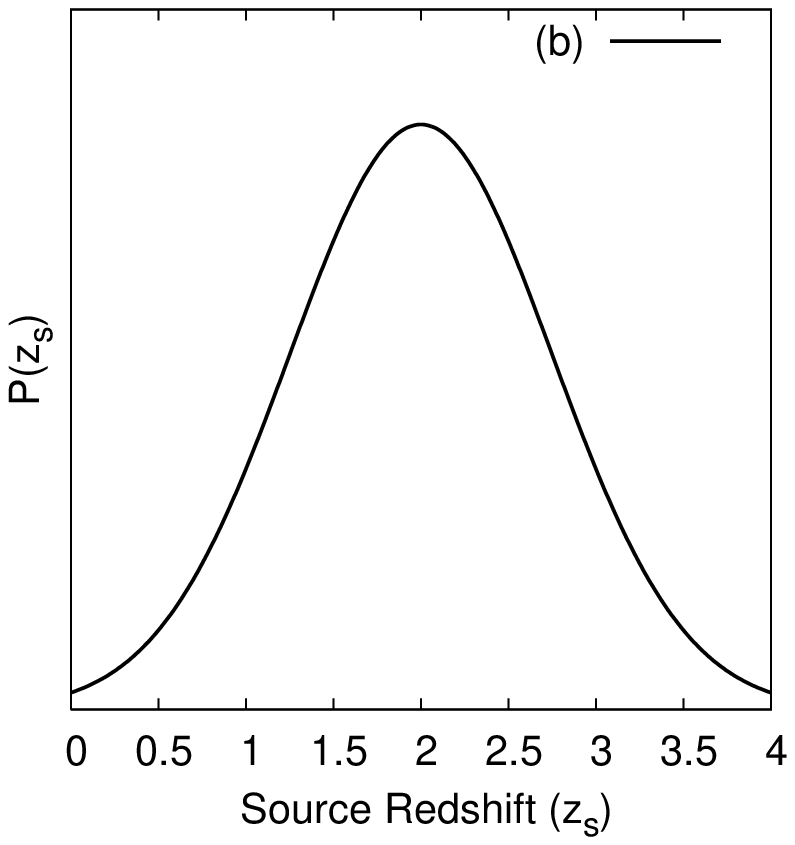,scale=0.448}
\epsfig{file= 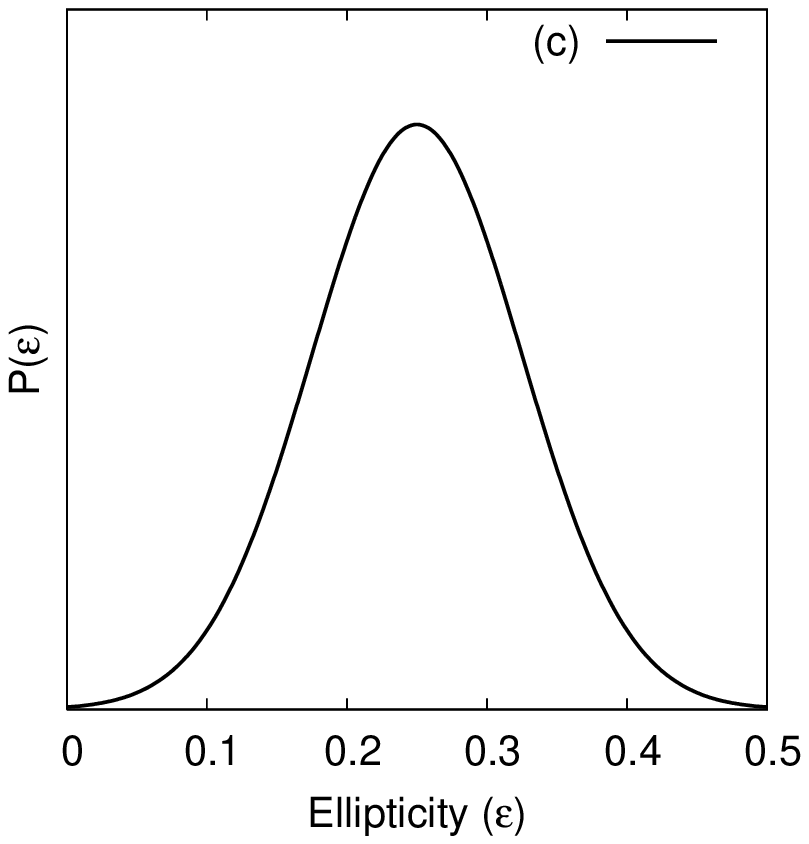,scale=0.448}
\epsfig{file= 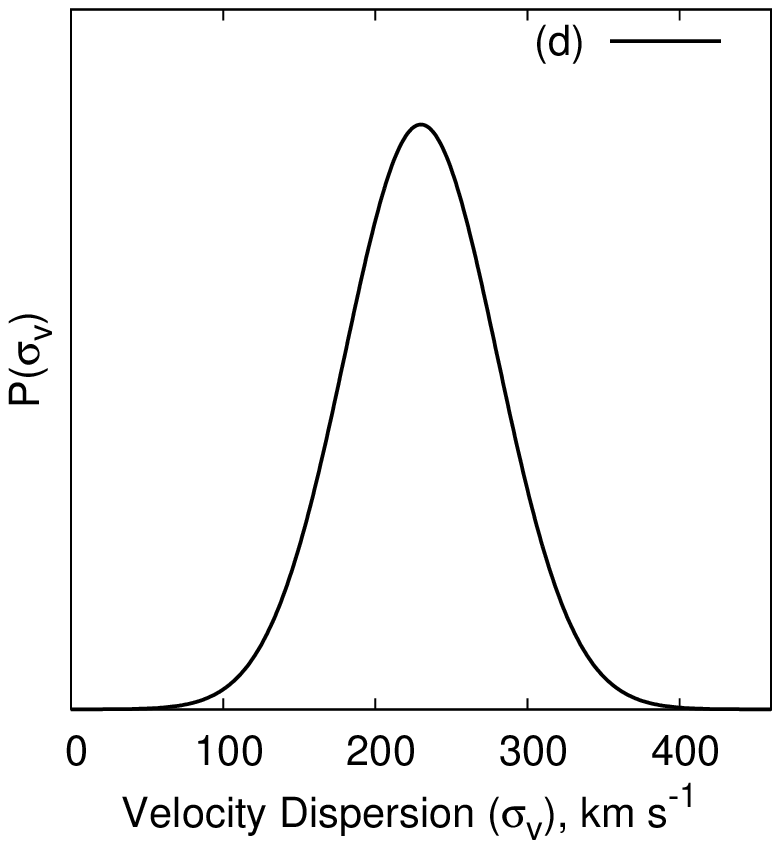,scale=0.448}
\epsfig{file= 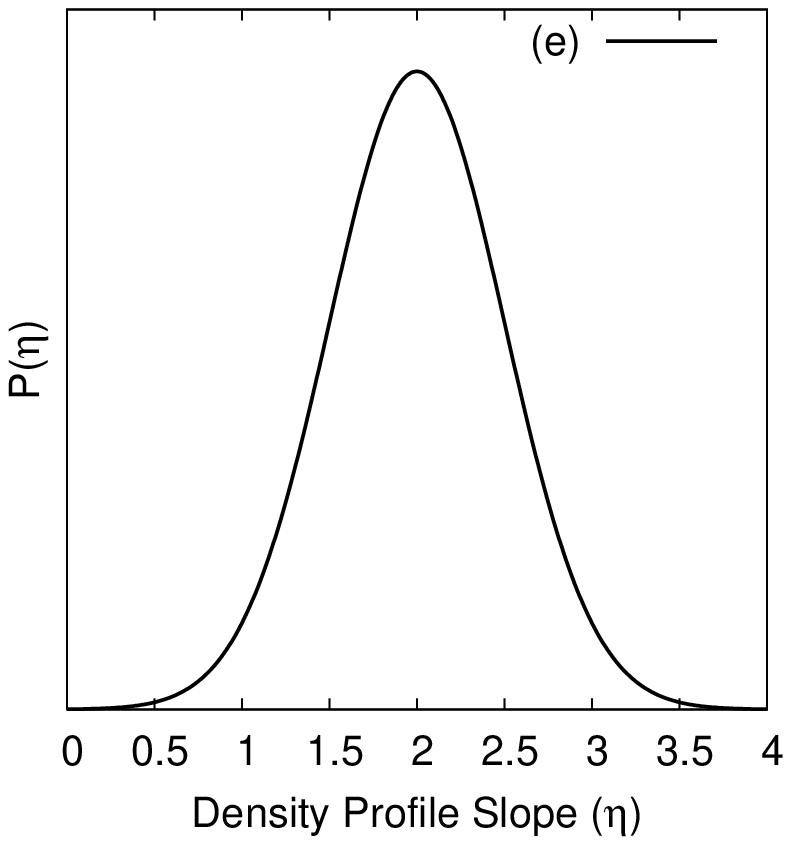,scale=0.448}
\epsfig{file= 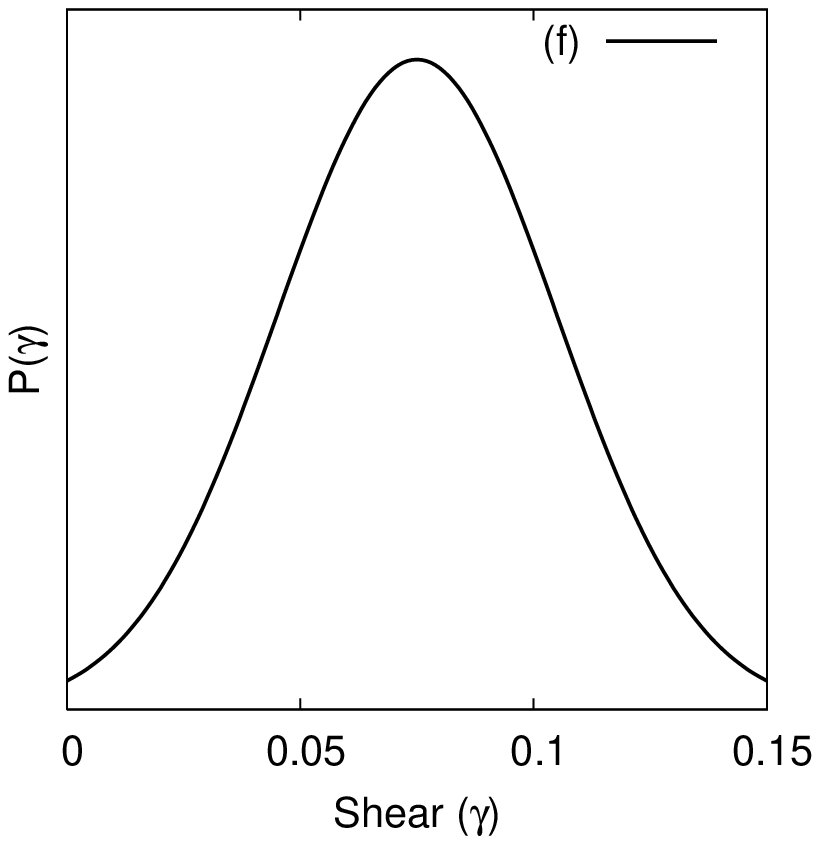,scale=0.448}
\epsfig{file= 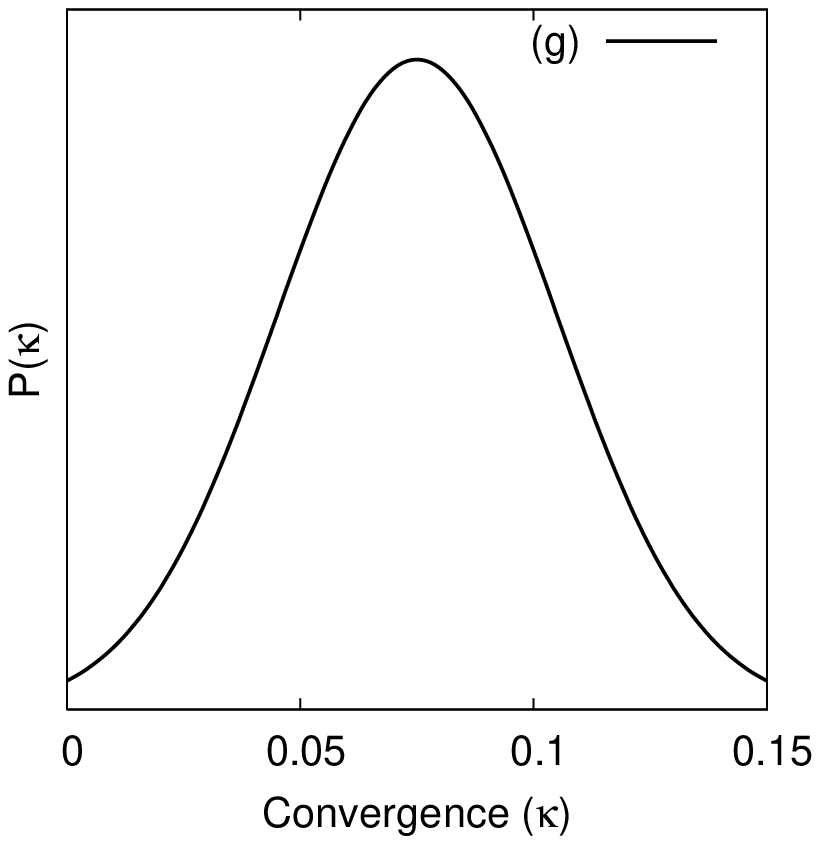,scale=0.448}
\caption[Estimates of predicted parameter distributions for future lens samples]{Estimates of predicted parameter distributions for future lens samples.  Gaussian properties from top left to bottom right:  (a) Lens redshift, $\mu$ = 0.5, $\sigma$ = 0.15, (e.g \citealt{wein}; \citealt{lapp}) ; (b) Source redshift,  $\mu$ = 2.00, $\sigma$ = 0.75, (e.g. \citealt{schn}; \citealt{tang}) ; (c) Ellipticity, $\mu$ = 0.25, $\sigma$ = 0.1, (\citealt{bak}; \citealt{khai}) ; (d) Velocity dispersion, $\mu$ = 230 km\,s$^{-1}$, $\sigma$ = 50 km\,s$^{-1}$, (e.g. \citealt{shel}) ; (e) Density profile slope, $\mu$ = 2.00, $\sigma$ = 0.50, (e.g. \citealt{rusi}; \citealt{koop1}) ; (f) Group shear, $\mu$ = 0.075, $\sigma$ = 0.025, (e.g. \citealt{dala}; \citealt{hold}) ; (g) Group convergence, $\mu$ = 0.075, $\sigma$ = 0.025, (e.g. \citealt{dala}; \citealt{hold})}
\label{plot1}
\end{figure*}

We begin with the generation of a synthetic data set upon which an analysis can be performed.  To accomplish this it is important to put into place a justifiable prescription for predicted features of future lens samples, specifically what characteristic value parameters that define the sample may take. For simplicity, below and in the analysis that follows, we assume Gaussian distributions for quantities describing the lens and source population characteristics (lens and source redshifts, lens galaxy ellipticity, velocity dispersion and slope of surface mass density profile, and convergence and shear of lens galaxy environment).  In addition, we do not account for covariance between parameters. Clearly, while Gaussian distributions and neglect of covariance serves as a fairly good approximation to some of these quantities, for others this is a very crude representation. The implications for these approximations are discussed in \S5.

The key parameters that will govern our sample are as follows:

\noindent \emph{1) Lens and source redshift ($z_{l}$, $z_{s}$)};  We currently have good constraints on source quasar redshifts from surveys such as SDSS (e.g. \citealt{schn}) and the Two-Degree Field QSO Redshift Survey (2QZ) (e.g. \citealt{tang}), which include sample sizes in excess of 40,000 QSOs.  Similarly, since we know that the vast majority of lens galaxies are early-types, we can again make use of large studies such as the DEEP survey (e.g \citealt{wein}) and the ESO-Sculptor survey (e.g. \citealt{lapp}), both of which consist of $\sim 500$ early-type galaxies (note that SDSS is too shallow for this purpose).  These samples indicate a peak in the lens and source redshifts of around $z_{l}$ = 0.4-0.6 and $z_{s}$ = 1.70. However, we also need to take into consideration the greater sensitivity of future observatories.  For example, predictions for the peak lens and source redshifts from both the SNAP mission concept and LSST \footnote{See http://www.lsst.org} are closer to $z_{l}$ $\sim$ 0.5 and $z_{s}$ $\sim$ 2.0 (\citealt{mars}).  In addition, the existing sample of lenses have lens and source redshift distributions more in line with these slightly higher estimates 
\footnote{See castles lens database; http://www.cfa.harvard.edu/castles/}.  The input redshift distributions for both lenses and sources are estimated from the above data, with widths of $\sigma$ = 0.15 and $\sigma$ = 0.75 respectively,  and are shown in Fig.\,\ref{plot1}a and \ref{plot1}b.

\noindent \emph{2) Galaxy ellipticity ($\epsilon$)};  As stated, the majority of lens galaxies exhibit some form of ellipticity.  We input an estimate of the predicted distributions in ellipticity on the sky as taken from \citet{bak} and \citet{khai}, where they measure the distributions of almost 1800 elliptical galaxies giving a mean of $\epsilon$ $\sim$ 0.25 (as defined via the axis ratio of the light; $\epsilon$ = 1 - b/a).  This is a very similar level of ellipticity as found in the Sloan Lens ACS (SLACS) survey sample ($\epsilon$ $\sim$ 0.24) as measured in both the mass via strong lens modelling and in the light, although this is from a sample size of just 15 \citep{treu}.  The input distribution used in the generated data set, with corresponding width of $\sigma$ = 0.1, is displayed in Fig.\,\ref{plot1}c.

\noindent \emph{3) Velocity dispersion ($\sigma_{v}$)};  Once again, we use data from the SDSS survey taking velocity dispersion distributions from a sample of almost 50,000 early-type elliptical galaxies \citep{shel}.  Here they split the available sample in half, giving velocity dispersion distributions for both low and high $\sigma_{v}$ ellipticals, the means of which were $\sigma_{v}$ = 143 km\,s$^{-1}$ and $\sigma_{v}$ = 215 km\,s$^{-1}$ respectively.  Lenses will have a bias towards the high mass systems since the lensing cross-section scales with greater mass/velocity dispersion.  Hence the higher $\sigma_{v}$ sample represents a better estimate for the distribution of lens galaxy velocity dispersions.  Indeed, this is also in keeping with the SLACS sample of 15 lenses which had a peak velocity dispersion of between $\sim$ 240-280 km\,s$^{-1}$.  The greater sensitivity of future instruments may however enable the detection of lensing by slightly lower mass systems.  Hence we take an approximate mid-point between the SDSS and SLACS sample and use a mean of 230 km\,s$^{-1}$ in our synthetic sample.  The form of this distribution, with corresponding width of $\sigma$ = 50 km\,s$^{-1}$, is shown in Fig.\,\ref{plot1}d. 

\noindent \emph{4) Density profile slope ($\eta$)};  While often a contentious issue, it seems that most lens galaxies studied so far do indeed have a nearly isothermal mass distribution.  Studies such as \citet{rusi} and \citet{koop1} use lensing data (combined with stellar dynamics data in the latter work) from 
22 and 15 lens systems respectively to obtain values very near $\eta$ = 2 for a density distribution that follows a form $\rho_{\rm tot} \propto \rho_{0}$\,r$^{-\eta}$.  Indeed, it is likely that on average the slope will converge in a large sample (i.e $\geq$ 100) to a value which, on current indications, is around $\eta$ = 2. There are however lenses that do appear to have density profiles that are significantly steeper or shallower than isothermal (e.g. \citealt{kocha}; \citealt{treub}; \citealt{kochc}; \citealt{auge3}; Read, Saha \& Macci\`{o} 2007) for a number of possible reasons (e.g. \citealt{dobkeb}; \citealt{auge}; \citealt{rea}).  In order to accommodate such lenses we introduce a significant spread to the distribution of $\sigma$ = 0.50 around our isothermal mean of $\eta$ = 2.  This is shown in Fig.\,\ref{plot1}e.  We will discuss the factors effecting the density profile and its degeneracies further in \S2.1.

\noindent \emph{5) Group convergence and shear ($\kappa$, $\gamma$)};  It is harder to estimate the average contributions from the group environment, particularly from observational data.  For this study we have chosen to consider expected distributions as calculated from N-body simulations of group galaxies.  In \citet{dala}, both the group convergence and shear are found at levels of $\sim$ 0.03 - 0.05, whereas \citet{hold} find external shear values closer to $\sim$ 0.11.  Again, lenses may be biased toward systems with a higher group convergence and hence may be more likely to be found in future surveys (e.g. \citealt{momc}; \citealt{fass}; \citealt{auge2}).  Similarly, shear values may be correspondingly higher -- some modelling examples of the time delay lens PG1115+080 require a \emph{minimum} of $\gamma$ = 0.05 to successfully reproduce image positions \citep{witt}.  Given these factors, for this study we have chosen to increase the mean group convergence over those found in the N-body investigations to $\kappa$ = 0.075, and to select the mid-point of the shear from the two studies, i.e. $\gamma$ = 0.075.  We display our input distributions in Figs.\,\ref{plot1}f and \ref{plot1}g respectively.  Both have corresponding widths of  $\sigma$ = 0.025. 

We note that all the distributions given above and shown in Fig.\,\ref{plot1} are Gaussian fits using the available surveys and studies, and allow us to generate sample data.  While it is not expected that similar Gaussian fits to future samples will change significantly from the distributions given, variation is inevitable.  To help compensate for this effect, we have generally increased the spread of the distributions over that taken from the literature, as highlighted earlier for the case of the density profile slope.

Once we have obtained reasonable estimates of parameter distributions, we can begin constructing synthetic data sets.  Unlike the distributions discussed above, the source position ($\beta_{1}$, $\beta_{2}$) for each system is selected from a uniform distribution.  For the purpose of this investigation, the data sets will consist of the following observables -- image positions and the corresponding time delays, as generated from a particular set of input parameters; velocity dispersion ($\sigma_{v}$), density profile slope ($\eta$), ellipticity ($\epsilon$), source position ($\beta_{1}$, $\beta_{2}$), source and lens redshift ($z_{s}$, $z_{l}$), the Hubble constant ($H_{0}$), external shear and convergence ($\gamma$, $\kappa$).  Additional constraints can be obtained with the inclusion of fluxes but traditionally it has been hard to make models fit well to the observations (e.g. due to micro-lensing, substructure effects etc.), so we exclude flux ratio constraints.  We use an elliptical power-law potential to generate time delay surfaces, then locate image positions and calculate the corresponding time delays.  The potential has the form

\begin{equation}
\psi (\theta_{1},\theta_{2}) = \frac{b^{2}}{(3-\eta)}\bigg[\frac{[(1-\epsilon)\theta^{2}_{1} + (1+\epsilon)\theta^{2}_{2}]^{1/2}}{b}\bigg]^{3-\eta}\,,
\label{equ:pl}
\end{equation}

\noindent where

\begin{equation}
b = 4 \pi \frac{\sigma_v^2}{c^2}\frac{D_{ls}}{D_s}\,,
\label{equ:lensstrength}
\end{equation}

\noindent with $D_{s}$ and $D_{ls}$ being the angular diameter distances between the observer and source, and the lens and source respectively.  The terms $\theta_{1}$ and $\theta_{2}$ are the angular image positions in the image plane.  We use an elliptical power-law potential model for speed and simplicity and for the reason that when investigating low ellipticity galaxies as is done in this study, the surface mass density corresponding to a power law potential model is virtually indistinguishable from that of a true surface mass density model \citep{kass}.  We model the contribution of the external group convergence and shear as follows (e.g. \citealt{keeta})

\begin{equation}
\psi_{\kappa,\gamma}(\theta_{1},\theta_{2}) = \frac{\theta_1^{2} + \theta_2^{2}}{2}[\kappa + \gamma cos 2(\theta - \theta_{\gamma})]\,,
\end{equation}

\noindent where the positions $\theta_{1}$ and $\theta_{2}$ are related to the position angle, $\theta$, via $\theta_{1}$ = $rcos\theta$ and $\theta_{2}$ = $rsin\theta$ with $r$ = $\sqrt{\theta_{1}^{2} + \theta_{2}^{2}}$, and $\theta_{\gamma}$ is the shear angle.

\subsection{Galaxy density profiles and associated degeneracies}

As highlighted above, it is difficult to simultaneously constrain the density profile slope (and its evolution) and $H_{0}$, due to their strong degeneracy.  We mentioned in our discussions above that certain time delay lenses appear to have steeper density profiles than those determined from the SLACS sample.  In principle this could result from a systematic slope offset between objects with well-measured slopes where time delay measurements are not possible, and lenses with time delays but poorly determined slopes.  We note, however, there are also time delay systems which yield $H_{0}$ $\ga$ 72 km s$^{-1}$ Mpc$^{-1}$ when modelled with an isothermal lens, implying a shallower density profile e.g. B0218+357 \citep{york}, B1608+656 \citep{koop3}, HE1104-1805 \citep{ofek}, Q0142-100 \citep{akhu}.  In addition, there is not any direct evidence that the SLACS sample is insensitive to such systematic errors, and therefore the SLACS distribution of $\eta$ may represent the true underlying distribution including any systematic effects.  When considering the analysis performed in \citet{ogur}, we also note that on an individual basis, before any combination of data was performed, the  $H_{0}$ values are scattered in an approximately symmetric way around $H_{0}$ = 72 km s$^{-1}$ Mpc$^{-1}$, and that this could be attributed to the approximately symmetric nature of the systematic errors that effect mass distribution profiles.  Indeed, based on the best currently available empirical data, it seems that a broad Gaussian distribution based around a value of $\eta$ = 2 is a fair starting point for the analysis that follows.  Another factor is that the measured profile slope could possibly vary with radius.  This would be a concern if the slope is calculated by comparing lensing constraints to dynamical information (e.g., \citealt{treuc}).  The velocity dispersion measurements are made at $R$ $\la$ $R_{eff}$, while the lensing information comes from the Einstein ring radius, $R_{Ein}$.  An effective slope, $\eta_{eff}$, calculated between these radii may not  match the local value of $\eta$ in the annulus between the lensed images if there is a significant difference between $R_{eff}$ and $R_{Ein}$.  Derivations of the Hubble Constant are particularly sensitive to any variation in the local value of $\eta$, and to achieve percent level accuracy one would need any slope correlation to be approximately d$\eta$/dR $\la$ 0.02 kpc$^{-1}$ around the region of the Einstein radius.  However, it appears from the SLACS study at least, that there are no correlations between the density slope and the effective radius, the Einstein radius, the distance between the two radii, and the mean of the two radii \citep{koop1}.  These measurements are based on 15 lenses in the SLACS sample, and an analysis based on a larger sample of $\sim$100 systems will help to rule out any such correlation.


We note that there is also a possibility that galaxy density profiles could evolve with redshift/time, and could therefore be correlated with certain factors e.g. with the angular diameter distance to the lens.  The best observational evidence on evolution comes from combining the SLACS and LSD (Lenses Structure and Dynamics; \citealt{koop4}) data sets; assuming a simple linear dependence of $\eta$ on $z$ gives d$\eta$/dz = 0.23$\pm$0.16 (1$\sigma$) over the range $z= 0.08 - 1.01$, marginally consistent with no evolution.  However, the mean of the Lenses Structure and Dynamics (LSD; \citealt{koop4}) sample is outside the joint-analysis range from the SLACS lenses, which could either be a redshift effect, since the LSD lenses are at a higher redshift range than those of the SLACS study, or simply be due to the small number statistics involved in the LSD study.  To beat down the $\pm$0.16 error to a level of around $\pm$0.02 would require approximately 80 additional lens systems.  More important however is the redshift coverage of that sample.  Due to the limited depth of SDSS much of the SLACS sample used in the \cite{koop1} study is only out to a redshift of z  $\sim$ 0.3.  A deeper sample of lenses, as will be discovered in the next generation of lens surveys, together with an even coverage over the z = 0-1 range, will help us understand the true evolution of slope vs. redshift.

The SLACS study now has around 80 lens systems and dynamical mass estimates of higher redshift systems are being determined - this will help in better understanding the evolution of the density profile (\citealt{bolt1}; \citealt{bolt2}).  Additionally, analysis of large N-body simulations may help to establish whether any subtle evolution should be expected.  Regardless, the evolution certainly does not appear to be extreme and for this study we assume that there is no correlation between redshift and density slope.  In the future, with large increases in lens survey data, stronger constraints on evolution will be derived and incorporated into analyses.

The steepness of the lens density profile is not the only factor that can be degenerate with the Hubble constant.  Shape degeneracies can also exist, for example in the form of varying and twisting ellipticities (see \citet{sahab} for discussions).  Recent studies have shown that the projection of intrinsically 
triaxial lens galaxies cannot effect the derivation of the Hubble constant \citep{corl}.  Further studies are required in order to explore other possible shape degeneracies.  For now, however, it is not entirely clear that the SLACS lenses would not have fairly sampled this distribution and at present the best empirical distribution of lens galaxy density slopes is approximately Gaussian around $\eta$ = 2.

Given larger upcoming samples, it may well be feasible to define a pre-selection process for systems that are to be included when deriving constraints on cosmological parameters.  For example, systems in environments with a richness above a certain threshold, knowingly prone to profiles that deviate from isothermality (e.g. \citealt{auge}) could be excluded.  It is possible that these selection criteria could introduce other biases into the sample.  On the other hand, should no measurable differences in results be found between the inclusion and exclusion of such systems, their relative importance could be better understood.

\section{Model and analytic method}

\subsection{The angular diameter distances}

The dependance on cosmology in this study comes via the observer-lens, observer-source and lens-source angular diameter distance expressions, $D_{l}$, $D_{s}$ and $D_{ls}$ respectively. Angular diameter distances are related to line-of-sight comoving distances

\begin{equation}
D_{c} = \frac{c}{H_{0}}\int^{z}_{0} \frac{dz'}{E{(z')}}\,,
\end{equation}

\noindent where the function $E$($z$$'$) depends on $\Omega_{\rm m}$, $\Omega_{\Lambda}$, and $\Omega_{K}$ in the following way

\begin{equation}
E(z) = \sqrt{\Omega_{\rm m}(1+z)^{3} + \Omega_{K}(1 + z)^{2} + \Omega_{\Lambda}}\,.
\end{equation}

$D_{l}$, $D_{s}$, $D_{ls}$ depend on $D_{c}$ in different ways.  When $\Omega_{K}$ = 0, as will be assumed in the following analysis, consistent with observations of the CMB e.g. \cite{sper}, we have

\begin{equation}
D_{l}(z) =  \frac{D_{c}(z_{l})}{1 + z_{l}}\,,
\end{equation}

\begin{equation}
D_{s}(z) =  \frac{D_{c}(z_{s})}{1 + z_{s}}\,,
\end{equation}

\begin{equation}
D_{ls}(z) =  \frac{1}{1 + z_{s}}|D_{c}(z_{s}) - D_{c}(z_{l})|\,.
\end{equation}

\subsection{Elliptical power-law lens model}

We model the data set described in \S2 with an elliptical power-law (EPL) model, the form of the potential being that of expression (\ref{equ:pl}).  The corresponding two dimensional lens equation is given by the lens equation $\vec{\beta}$ = $\vec{\theta}$ -- $\nabla\psi(\vec{\theta})$, where $\psi(\vec{\theta})$ is the combination of the lens potential and the potential contribution from the external convergence and shear, i.e. $\psi_{tot}$ = $\psi_{epl}$ + $\psi_{\kappa,\gamma}$.  Hence for the two components of $\beta$ we have

\begin{eqnarray}
&&\nonumber \beta_{1} = (1 - \kappa + \gamma) \theta_{1} \\
&& \,\,\,\,\,\,\,\,- \bigg[\frac{b(1-\epsilon)\theta_{1}([(1-\epsilon)\theta^{2}_{1} + (1+\epsilon)\theta^{2}_{2}]^{1/2}/b)^{2-\eta}}{[(1-\epsilon)\theta^{2}_{1} + (1+\epsilon)\theta^{2}_{2}]^{1/2}}\bigg]\,,
\label{equ:le1}
\end{eqnarray}

\noindent and

\begin{eqnarray}
&& \nonumber \beta_{2} = (1 - \kappa - \gamma) \theta_{2} \\
&& \,\,\,\,\,\,\,\,- \bigg[\frac{b(1+\epsilon)\theta_{2}([(1-\epsilon)\theta^{2}_{1} + (1+\epsilon)\theta^{2}_{2}]^{1/2}/b)^{2-\eta}}{[(1-\epsilon)\theta^{2}_{1} + (1+\epsilon)\theta^{2}_{2}]^{1/2}}\bigg] .
\label{equ:le2}
\end{eqnarray}

The time delay at a particular image position (with respect to the light travel time of the unperturbed ray in the absence of lensing) calculated by the solution of the lens equations has the form (see e.g. \citealt{nara})

\begin{equation}
t(\vec{\theta}) = \bigg[\frac{1+z_{l}}{c}\bigg]\bigg[\frac{D_{l} D_{s}}{D_{ls}}\bigg]
\bigg[
\frac{1}{2} \left|\vec{\theta}- \vec{\beta}\right|^{2} - \psi_{tot}(\vec{\theta})\bigg]\,.
\label{equ:time} 
\end{equation}

The resultant time delay $\Delta t$ between two image positions is therefore the corresponding difference in the light-travel time given by (\ref{equ:time}).





The lens equations (\ref{equ:le1}) and (\ref{equ:le2}) are solved simultaneously using the Newton-Raphson method for a non-linear system of equations.  We employ the MNEWT routine \citep{pres} which requires the two lens equation components and the corresponding Jacobian matrix.  This routine returns the image positions, $\vec{\theta_{i}}$, that satisfy the lens equations for a given set of free parameters; $\sigma_{v}$, $\eta$, $\epsilon$, $\beta_{1}$, $\beta_{2}$, $H_{0}$, $\gamma$, $\kappa$.  For a given set of image solutions and parameters the model time delays are then calculated using (\ref{equ:time}).  The image positions ($\theta_{i}$) and corresponding time delays ($t_{j}$) are then compared to those from the synthetic data set of $\S$2 using $\chi^{2}$ statistics,



\begin{eqnarray}
&&\nonumber \chi^{2} = \displaystyle\sum_{i=1}^n\frac{(\vec{\theta}_{i,Mod} - \vec{\theta}_{i,Obs,\theta})^{2}}{\sigma_{\theta(Obs)}^{2}} \\
&&\,\,\,\,\,\hspace{0.2cm} + \displaystyle\sum_{j=1}^n\frac{(\Delta t_{j, Mod} - \Delta t_{j,Obs})^{2}}{\sigma_{\Delta t(Obs)}^{2}}\,,
\label{equ:chi}
\end{eqnarray}

\noindent where $\sigma_{\theta(Obs)}$ and $\sigma_{\Delta t(Obs)}$ are the errors in the synthetic observed positions and time delays which are taken as conservative estimates of 1\% and 10\% respectively.  We minimise expression (\ref{equ:chi}) by employing the POWELL minimisation routine \citep{pres} which returns the best fit parameters for any particular system in an array comprising [$\sigma_{v}$, $\eta$, $\epsilon$, $\beta_{1}$, $\beta_{2}$, $H_{0}$, $\gamma$, $\kappa$], for a given choice of $\Omega_{\rm m}$ and $\Omega_{\Lambda}$, which is explored on a grid.  As noted above, we assume a flat Universe with $\Omega_{K}$ = 0.  The corresponding $\chi^{2}$ grids are combined from each system, allowing for the investigation into how the constraints scale with the number of  systems combined for any given survey data set.

\begin{figure*}
\centering
\epsfig{file= 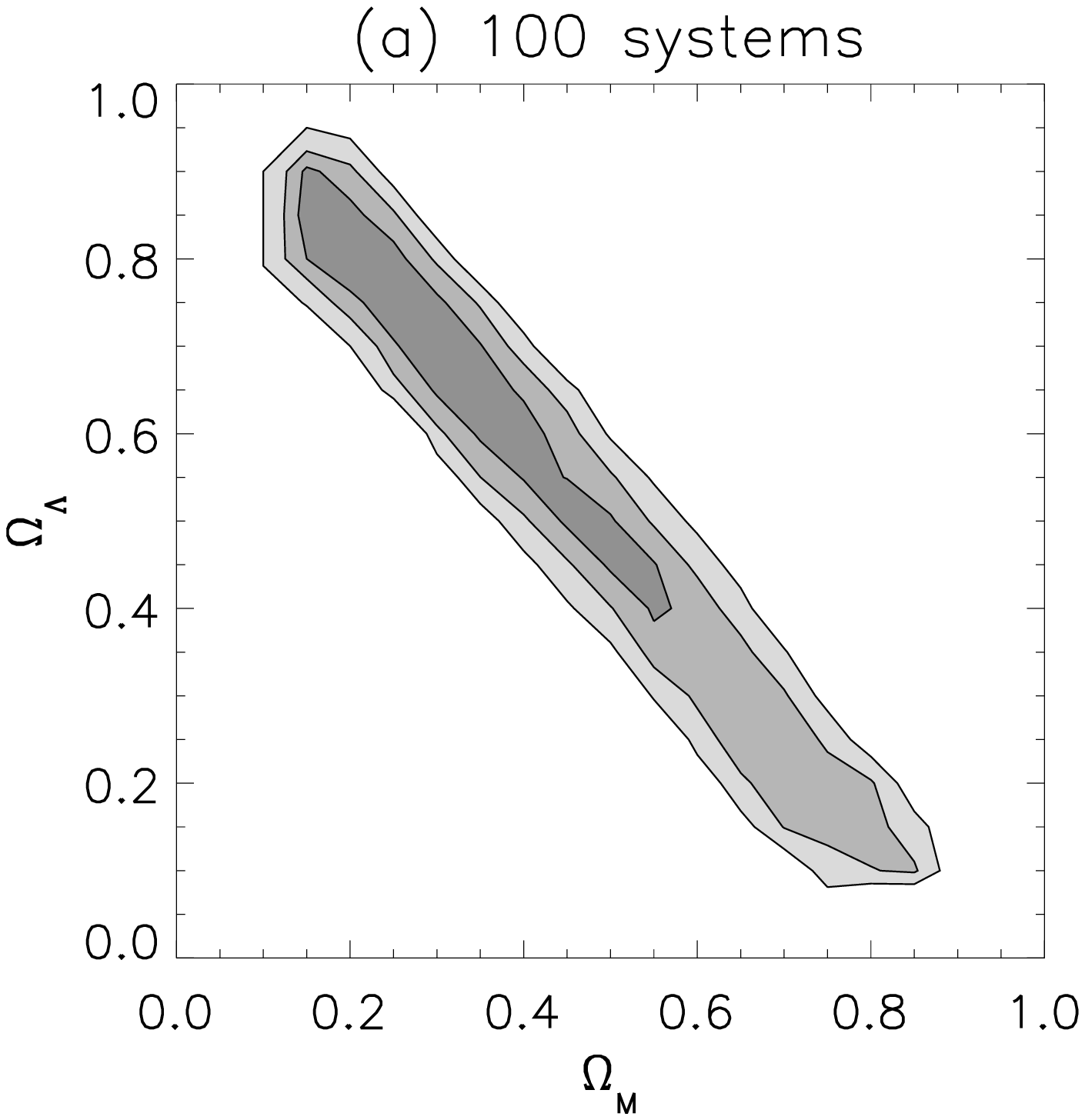,scale=0.39}
\epsfig{file= 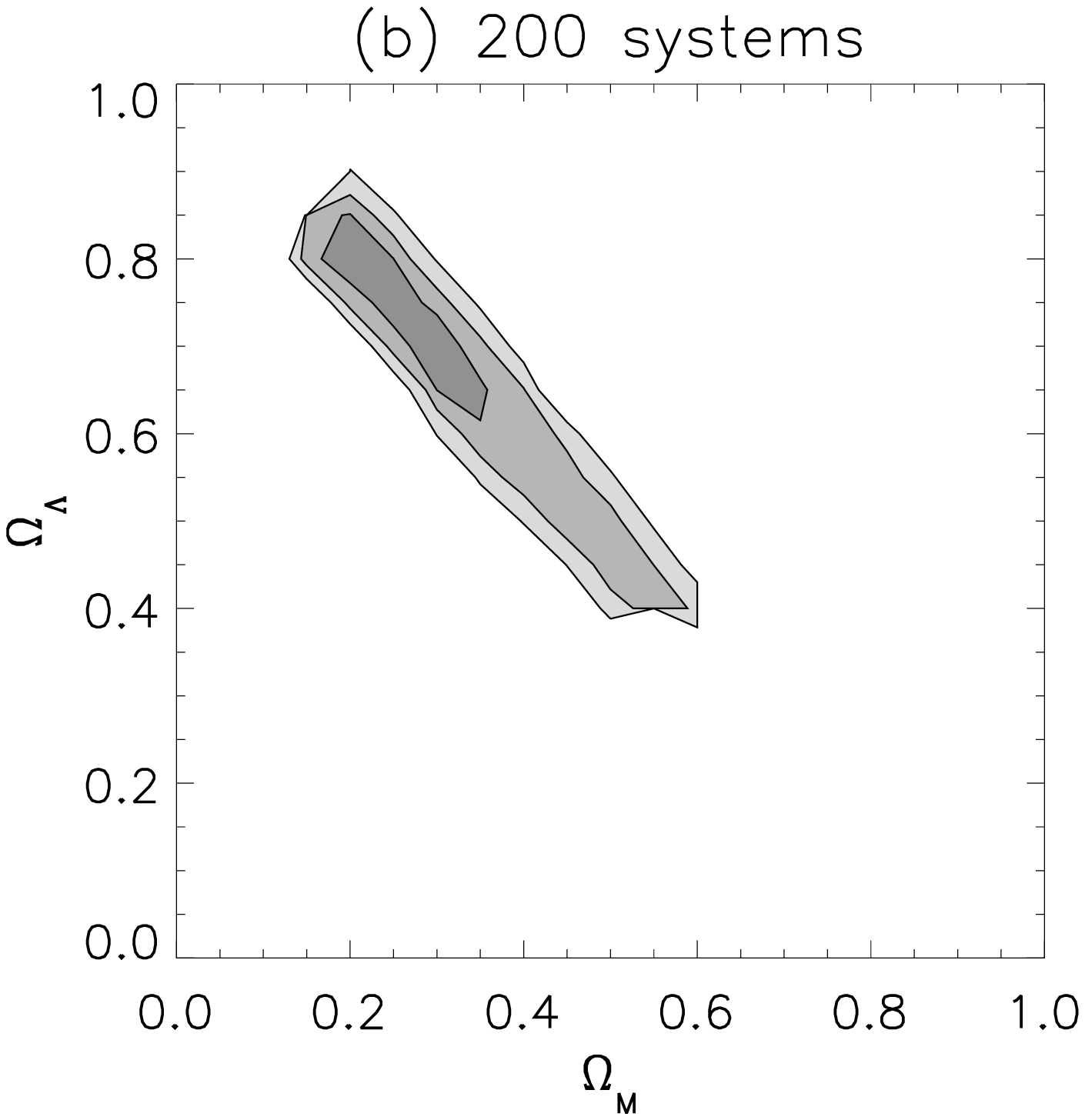,scale=0.39}
\epsfig{file= 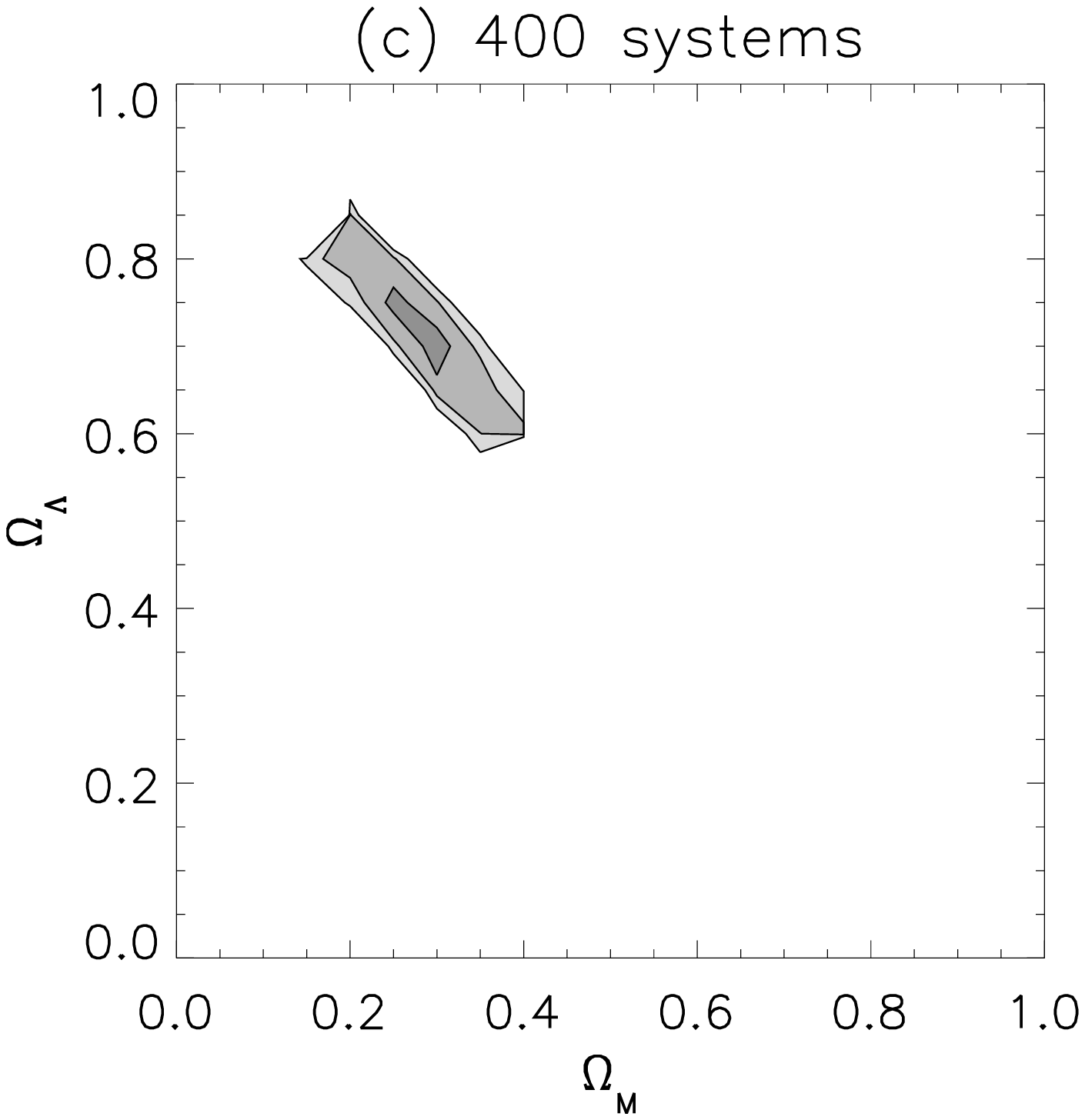,scale=0.39}
\caption[Constraints on the $\Omega_{\rm m}$ - $\Omega_{\Lambda}$ plane with increasing number of time delay lens systems]{
(a) The $\Omega_{\rm m}$ -- $\Omega_{\Lambda}$ plane as constrained from a synthetic data set consisting of 100 time delay lens systems.  Plotted are the 1, 2, and 3$\sigma$ contours.  We see a strong degeneracy along the $\Omega_{\rm m}$ + $\Omega_{\Lambda}$ = 1 line.  Constraints are approximately similar in scale to those from e.g. early CMB experiments, SN1a measurements, X-ray clusters, but in comparison to more modern day instruments it provides us with no new limits ;  (b)  Same as left plot but for 200 systems.  The degeneracy is still present but the contours have shrunk considerably, particularly along the $\Omega_{\rm m}$ + $\Omega_{\Lambda}$ = 1 direction ; (c)  Constraints are now based upon data from 400 systems.  There is significant closure in contour size when compared with 100 and 200 systems.  This level of parameter constraint is similar in precision to modern CMB instruments, and it is now possible to recover the input cosmology ($\Omega_{\rm m}$ = 0.3, $\Omega_{\Lambda}$ = 0.7) to relatively high precision.}
\label{plot2}
\end{figure*}

\subsection{Model priors}

As highlighted in \S2, there is a great deal of information that we already know \emph{a priori} concerning the properties of early-type lens galaxies.  In any future analysis of a lens sample we would not necessarily know the precise values for parameters such as the slope of the density profile, or ellipticity, but it is the intention of this work to highlight that the \emph{bulk} properties of the sample are quite well defined, and as the sample size increases, the necessity to know the precise values for parameters potentially diminishes accordingly.  More important is the information about the sample as a whole, and what prior knowledge we can put into the prescription to help improve our constraints on cosmological parameters from lens samples.  To this end we place some additional Gaussian priors into the $\chi^{2}$ prescription given above (\ref{equ:chi}).  These include priors on e.g. the Hubble constant from the HST Key Project (\citealt{free}) when constraining the $\Omega_{\rm m}$ -- $\Omega_{\Lambda}$ plane.  Information on parameters such as velocity dispersion are essentially similar to those given in \S2, however since the exact distribution of the lens sample would not be known in reality, but rather only an estimate of the bulk properties, we expand those in \S2 and use far broader distributions as the actual priors.  The values used for the mean and variance of each prior on $H_{0}$, $\sigma_{v}$, $\epsilon$, $\eta$, $\kappa$ and $\gamma$ is displayed in Table 1, and the prior takes the form;

\begin{equation}
\hspace{3cm}\chi^{2}_{p} = \frac{(p - \mu_{p})^{2}}{\sigma_{p}^2}\,.
\label{equ:hub_prior}
\end{equation}

We note that when fitting two image systems, no prior on $\gamma$ is employed since such systems typically have low levels of external shear.

%





\section{Analysis and Results}

Below we employ our analysis method detailed in the previous section with the synthetically generated lens sample.  The input cosmology that is used to generate the lens sample is a $\Lambda$CDM model with $\Omega_{\rm m}$ = 0.3, $\Omega_{\Lambda}$ = 0.7, and h = 0.72.

\begin{table}
\caption{Model prior values as used in expression (\ref{equ:hub_prior})}
{\footnotesize
\begin{tabular}{cll}\hline\hline
Parameter ($p$) & Mean ($\mu_{p}$)& Variance ($\sigma_{p}$) \\
\hline
&& \\
\,\,\,\,\,\,$H_{0}$ & \,\,\,\,\,\,72.0 & \,\,\,\,\,\,16.0 \\ 
\,\,\,\,\,\,$\sigma_{v}$ & \,\,\,\,\,\,230 & \,\,\,\,\,\,200 \\
\,\,\,\,\,\,$\epsilon$ & \,\,\,\,\,\,0.25 & \,\,\,\,\,\,0.25 \\
\,\,\,\,\,\,$\eta$ & \,\,\,\,\,\,2.00 & \,\,\,\,\,\,2.00 \\
\,\,\,\,\,\,$\kappa$ & \,\,\,\,\,\,0.075 & \,\,\,\,\,\,0.15 \\
\,\,\,\,\,\,$\gamma$ & \,\,\,\,\,\,0.075 & \,\,\,\,\,\,0.15 \\
\hline 
\end{tabular}
}

\end{table} 

\subsection{Constraints on the \bf{$\Omega_{\rm m}$ -- $\Omega_{\Lambda}$} plane}

We incorporate the priors detailed in Table 1 and expression (\ref{equ:hub_prior}) into our $\chi^{2}$ function (\ref{equ:chi}).  We begin by analysing the $\Omega_{\rm m}$ -- $\Omega_{\Lambda}$ plane and examine the relationship between lens sample size and parameter constraints with samples of 100, 200, and 400 lens systems.  The results are shown in Fig.\,\ref{plot2}.

To begin with, Fig.\,\ref{plot2}a shows constraints based upon the analysis of 100 systems.  The predominant degeneracy lies along the $\Omega_{\rm m}$ + $\Omega_{\Lambda}$ = 1 direction.  Constraints at this level are approximately similar in scale to those from early CMB experiments (e.g. Maxima, COBE; \citealt{jaff}; \citealt{stom}), SN1a measurements and X-ray clusters (\citealt{alle}; \citealt{tonr}), but in comparison to modern day CMB instruments it provides us with no new limits.  In Fig.\,\ref{plot2}b we increase the sample size to 200 systems and see a notable reduction in the 2, and 3$\sigma$ contours.  This level of constraint in the $\Omega_{\rm m}$ -- $\Omega_{\Lambda}$ plane is similar to that from studies involving more modern CMB measurements (e.g. WMAP - DR1; \citealt{sper2}).  The  first data release of the WMAP instrument yielded the value $\Omega_{\rm m}$ = 0.29 $\pm$ 0.07 for WMAP data alone, compared to an error of $\pm$ 0.1 on the input value of $\Omega_{\rm m}$ = 0.30 for Fig.\,\ref{plot2}b above.  Similar error comparisons are found for $\Omega_{\Lambda}$ though not explicitly stated in DR1.

\begin{figure*}
\centering
\epsfig{file= 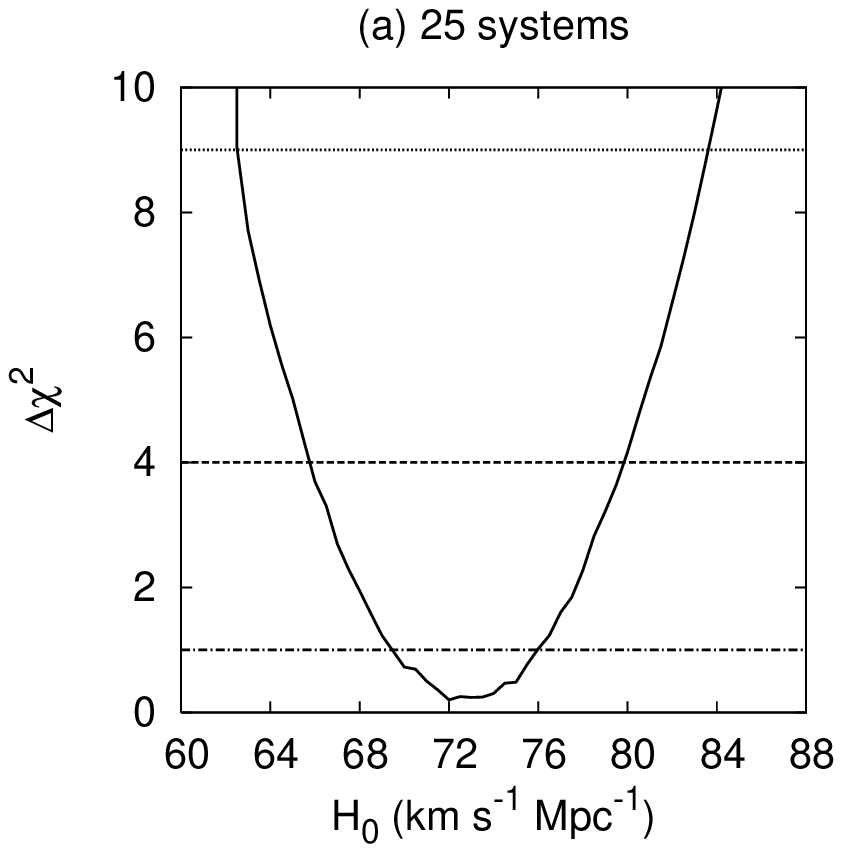,scale=0.65}
\epsfig{file= 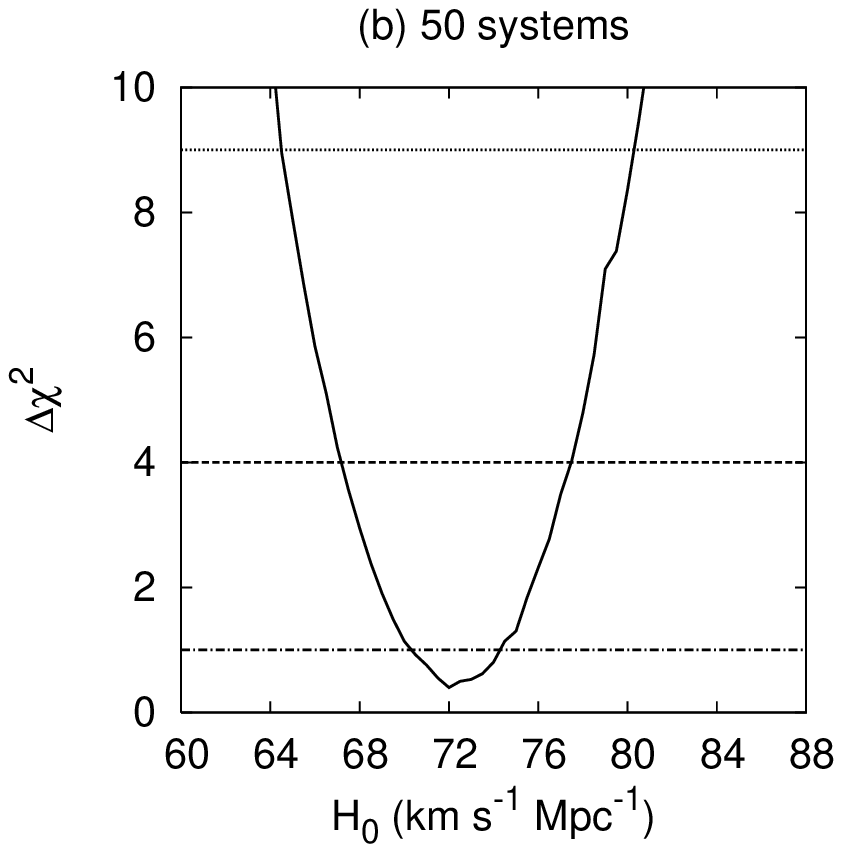,scale=0.65}
\epsfig{file= 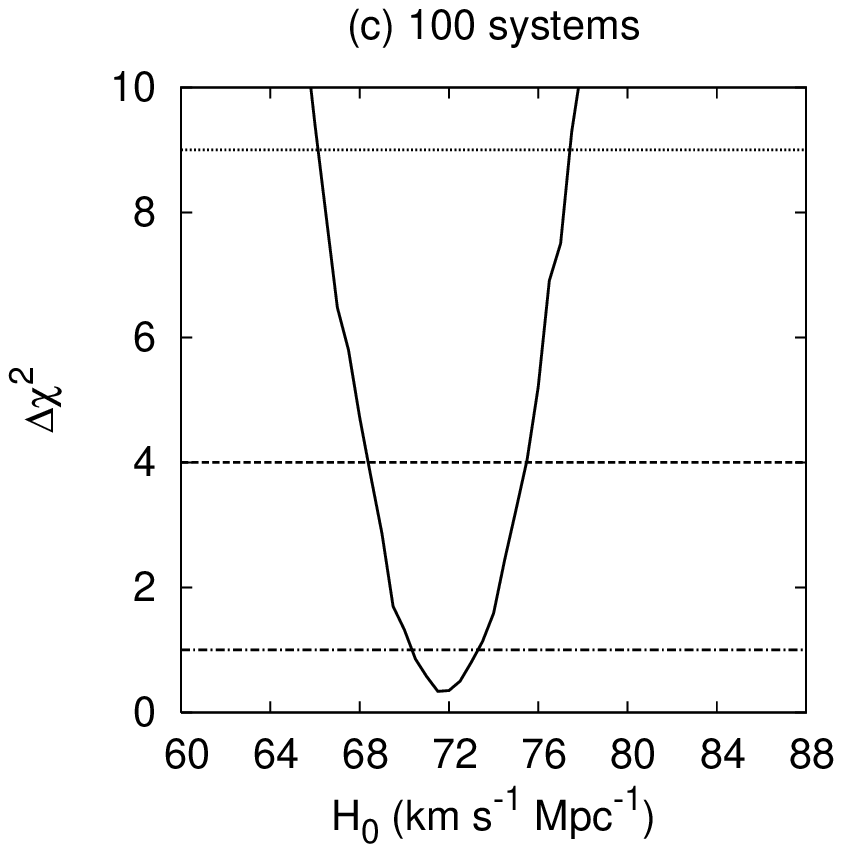,scale=0.65}
\caption[Variation of $H_{0}$ constraints with increasing number of lens systems]{(a) The constraints on $H_{0}$ based upon 25 time delay lens systems.  Horizontal lines mark the formal 1, 2, and 3$\sigma$ limits (dot-dash, dash and dotted respectively).  Sample size and the constraints obtained are similar in magnitude to those obtained from existing time delay samples ($\sim$20 systems, c.f. \citealt{ogur}) ; (b)  The sample size is now doubled to 50 time delay lenses showing the corresponding reduction in width ; (c) Constraints are now based upon 100 time delay systems, yielding 1$\sigma$ errors in the region of $\sim$1--2 \% uncertainty in $H_{0}$.
}
\label{plot3}
\end{figure*}

To examine the possible limit of this approach we double the number of systems again; Fig. \,\ref{plot2}c shows constraints based upon a sample of 400 lens systems with their corresponding time delays.  We now see a significant closure of all contours and recover the input cosmology ($\Omega_{\rm m}$ = 0.3, $\Omega_{\Lambda}$ = 0.7) with high precision.  For this sample size the 1, 2, and 3$\sigma$ errors are smaller than those produced by modern CMB experiments (e.g. WMAP - DR3; \citealt{sper}).  When assuming  \emph{w}  = -1 and the HST value for $H_{0}$, the third data release from the WMAP instrument yields values of  $\Omega_{\rm m}$ = 0.241 $\pm$ 0.034 and $\Omega_{\Lambda}$ = 0.716 $\pm$ 0.055.  In comparison, Fig. \,\ref{plot2}c above yields similar errors of $\pm$ 0.05 on the input values of $\Omega_{\rm m}$ = 0.30 and $\Omega_{\Lambda}$ = 0.70.  Naturally, future CMB measurement precision is also expected to improve, e.g. Planck (\citealt{xia}).  At this point, we note that even if ultimately the precision from this method is found to be less than that from CMB experiments, which is probable, tight constraints on parameters should be possible, and it is the complimentary nature of such a measurement that is valuable (see \citet{lind} for discussions).

So far we have assumed that the bulk properties of the data sample being analysed are well known i.e. that the mean of the priors match those of the synthetic data set distributions of \S2.  It is probable that this will not always be the case and that some prior knowledge is lacking with certain distributions.  To investigate the effects of this we re-run the analysis discussed above, but include a $\pm$1$\sigma$ shift in the mean of the priors (using the Gaussian widths of \S2, Figs.\,\ref{plot1}) to simulate a systematic under or over estimate in one of the bulk sample properties.  For the case when one out of the six priors in \S3.3 had their mean shifted, the recovered constraints remained the same or very similar as for those displayed in Fig.\,\ref{plot2}.  When we introduced systematic shifts in the means of two out of the six priors, the overall constraint was reduced to that typically obtained from half the original number of lens systems, e.g. a $\pm$1$\sigma$ shift in the means of both velocity dispersion and ellipticity resulted in constraints obtained for 400 lens systems that were similar in scale to those from 200 systems obtained with no shift in the means.  If a severe lack of prior knowledge was assumed in the analysis, and three or more of the six priors mismatched the synthetic sample, or if the shift was $\pm$2$\sigma$ or more, the constraints no longer minimised on the synthetic sample's starting cosmology of $\Omega_{\rm m}$ = 0.3, $\Omega_{\Lambda}$ = 0.7.  We highlight, that a $\pm$2$\sigma$ shift in e.g. the velocity dispersion distribution, would imply a severe lack of prior knowledge of a lens sample and is hard to foresee.  When small shifts of $<$ $\pm$1$\sigma$ where introduced, very little difference was observed in the recovered constraints, even when introduced into multiple priors.

The key investigation in this particular study, i.e. the constraint on cosmological parameters as a function of sample size, depended weakly on changes in the exact form of the initial starting distributions used to create the initial synthetic data set.  We tested this by creating different synthetic data sets after varying the form of the distributions discussed in \S2 (i.e. varying the quoted $\mu$ and $\sigma$ of Fig.\,\ref{plot1}), and repeating the analysis -- in all cases the basic relationship between the constraint on cosmological parameters as a function of sample size remained constant.  Far more important in the ability to obtain constraints on cosmological parameters via time delays is the prior knowledge of the bulk properties of that sample (whatever form they happen to initially take), in particular the prior knowledge on the mean values of the various distributions, as discussed above.

\subsection{Implications for $H_{0}$}

We have seen that with lens sample sizes of $\sim$ 400 time delay systems it is possible to obtain comparatively high precision constraints in the $\Omega_{\rm m}$ -- $\Omega_{\Lambda}$ plane.  As an independent method for the study of cosmological parameters this holds the potential to play a useful role in the future.  Larger sample sizes also have a direct impact on the study of the Hubble constant.  The current best value for the Hubble constant is still the HST Key Project value of 72 $\pm$ 8 km s$^{-1}$ Mpc$^{-1}$ \citep{free}.  A number of $H_{0}$ values are found from modelling individual lensing time delay systems and are in the range of about 50-80 km s$^{-1}$ Mpc$^{-1}$ (see e.g. \citealt{cour}; \citealt{kochb}; \citealt{jack}), although when considered as an ensemble good agreement with the HST value is found (e.g. \citealt{ogur}; \citealt{saha}).  The CMB does not directly measure $H_{0}$, but it is possible to infer a measurement (based on a number of analysis specific assumptions; see \citealt{sper}) of $H_{0}$ $\simeq$ 73 $\pm$ 3 km s$^{-1}$ Mpc$^{-1}$.  Future increases in supernova samples sizes (e.g. from SNAP and DESTINY) may help to increase the precision on $H_{0}$.  In this section, however, we focus on the possible improvements on precision that could one day come about due to large increases in the number of strong lensing time delays.

As highlighted in \S1, it is common practice to fix the cosmology to some pre-defined concordance cosmology as derived from a combination of measurements (e.g. SN1a \citealt{tonr}; CMB \citealt{sper}), and focus on obtaining estimates for $H_{0}$, and some success has been had with this approach.  One of the main obstacles to this method is the degeneracy that exists between $H_{0}$ and the slope of the density profile of the lensing galaxy.  As discussed however, at some particular sample size the average density profile should tend towards a predictable distribution that can help in breaking the degeneracy and from observations this value appears to be close to isothermal (e.g. \citealt{koop1}).  We explore the implications of this concept by investigating the variation of $H_{0}$ parameter constraint with increasing sample size, as done previously in the $\Omega_{\rm m}$ -- $\Omega_{\Lambda}$ plane.  We remove the HST prior, but keep the slope of the density profile as $\eta$ = 2 $\pm$ 1 and leave the other remaining priors as they are.  We display results in Fig.\,\ref{plot3}.  

We begin with a sample size of 25 systems (Fig.\,\ref{plot3}a).  Notice that the constraints placed on $H_{0}$ in this case are similar to those found with similar sized samples from existing observational data (e.g. \citealt{ogur}; \citealt{saha}), the current-day sample consisting of $\sim$ 20 time delay lenses, which put values at  70 $^{+9}_{-5}$ km s$^{-1}$ Mpc$^{-1}$ (2$\sigma$) and  72 $^{+8}_{-11}$ km s$^{-1}$ Mpc$^{-1}$ (2$\sigma$), respectively.

We double the sample size to 50 time delay lenses and show the results in Fig.\,\ref{plot3}b.  The contours shrink notably and we are now in the realm of precision that exceeds the HST Key Project value (72 $\pm$ 8 km s$^{-1}$ Mpc$^{-1}$).  Increasing the sample size to 100 systems moves beyond this level of precision, towards $\sim$1--2\% uncertainty in the recovered value (Fig.\,\ref{plot3}c).


As was done in the case for constraints on $\Omega_{\rm m}$ -- $\Omega_{\Lambda}$, shifts in the priors were introduced when considering constraints on $H_{0}$.  We noted similar broadening of the constraints over those shown in Fig.\,\ref{plot3} when introducing mean shifts in multiple priors, although in this case it had less of a negative effect, i.e. the width did not increase to the level found with half the number of systems as was the case for shifts discussed above for the $\Omega_{\rm m}$ -- $\Omega_{\Lambda}$ plane.  This is possibly due to $H_{0}$ being more sensitive to time delays than $\Omega_{\rm m}$ and $\Omega_{\Lambda}$, and hence having more of a capacity of overcoming poor prior knowledge.  Nevertheless, there was still an observable broadening of the contours and this should be noted.

\section{Discussion and Conclusions}

Future observatories are assured of finding a large number of strong gravitational lens systems, which have the potential of providing a unique and compelling probe of cosmology.  There are a number of studies that can be performed via this unique phenomena, one of which is the well documented use of constraining the value of the Hubble constant.  Less well explored, however, are the constraints that can be put on $\Omega_{\rm m}$ and $\Omega_{\Lambda}$ via lensing time delays, mainly due to the low numbers of such systems available for study at the present time.  Lens samples promise to grow rapidly once upcoming surveys get underway including the DES, LSST, JDEM, DUNE, and the SKA, which will  
uncover many thousands of strong lens systems. Investigations into what cosmological information these lens samples can yield could help shape the particular survey strategies themselves.

Lens statistics have already provided us with useful constraints in the $\Omega_{\rm m}$ -- $\Omega_{\Lambda}$ plane (e.g. \citealt{mitc}).  While some $\sim$ 100 multiply imaged quasars and radio sources have been discovered, a statistical analysis requires a sample from a survey that is complete and has well-characterised, homogeneous selection criteria. The largest such sample comes from CLASS (\citealt{brow}; \citealt{myer}), with a total of 22 confirmed lenses.  Future surveys, such as those performed by the LSST for example, will provide far greater sized statistical samples.  Indeed, we are already beginning to see the benefit of increased sample sizes for lens statistics from surveys such as SDSS \citep{ogur2}.

In this work we have demonstrated that usable cosmological constraints can be placed on the $\Omega_{\rm m}$ -- $\Omega_{\Lambda}$ plane with 100 time delay systems or more (c.f. $\sim$ 20 presently known time delay systems), a sample size that is likely to become available within several years.  When samples grow to several hundreds in the next decade, strong lensing has the potential to constrain such cosmological parameters to levels approximately equal to that of current CMB experiments, although of course these will have also increased in precision with future measurements. We note that analyses from lens statistics produce orthogonal constraints to those that are derived through time-delays (e.g. \citealt{mitc}; \citealt{ogur2}), revealing the possibility of dual constraints from the \emph{same statistical sample}.  This holds further potential in increasing the attainable precision for cosmology from strong lensing.

Currently the best estimate for the Hubble constant is from the HST Key Project at 72 $\pm$ 8 km\,s$^{-1}$ Mpc$^{-1}$ \citep{free}.  This still gives a rather broad 2$\sigma$ range of 56--88 km\,s$^{-1}$ Mpc $^{-1}$, a relatively large uncertainty for one of the most fundamental cosmological parameters that governs the length and time-scale of our Universe.  With the addition of new time delay systems from future surveys, lensing has the potential of providing a very high precision constraint on $H_{0}$.  As discussed previously, many of the existing lensing degeneracies, e.g. between the $H_{0}$ and the density profile slope, should become less of an issue with large samples and it is likely that on average the slope will converge in such a sample (i.e $\geq$ 100) to a value which, on current indications, is around $\eta$ = 2.  Indeed, we have seen in this study that larger samples hold the promise of reducing the uncertainty of the Hubble constant value to levels as low as 1--2\% uncertainty with the addition of a further 80 systems to the existing time delay sample.  In this study we have assumed a mean galaxy density profile that does not evolve with redshift, based on the current best estimate from the SLACS study, although there still remains an uncertainty in this matter.  The SLACS study now has over 80 lens systems and dynamical mass estimates of higher redshift systems are being determined - this may provide a better empirical understanding of mass slope evolution.  Large future surveys should be able to resolve this question.

With upcoming missions such as Planck and the SKA preparing to yield improved constraints on many cosmological parameters it is obvious to question why constraints from lensing should be pursued.  Lensing as a tool for cosmology is interesting for a key reason; one of the most vital questions in cosmology today is whether a dark energy component exists.  At present only the supernovae data set is able to address that question (CMB experiments can constrain the total density $\Omega_{\rm m}$ + $\Omega_{\Lambda}$, while clusters constrain $\Omega_{\rm m}$).  A significant result from lens statistics so far is the unique and independent evidence it provides for $\Omega_{\Lambda}$ $>$ 0, without any other cosmological assumptions (e.g. \citealt{mitc}).  The exact physics that allows Type Ia supernovae to be used as cosmological probes is still an area of uncertainty, and as such the information and confirmation obtained from lensing is highly valuable.  Indeed, it is the complimentary nature of constraints from strong lensing that could prove to be valuable, as discussed in \citet{lind}. 

In addition, other interesting possibilities exist to do cosmography with larger strong lensing samples, such as using samples of double Einstein rings, as recently discovered in the SLACS data set \citep{gava}, or by combining strong lensing and stellar dynamics data as suggested by \cite{gril}. While here we have focused on $w=-1$, strong lensing time delays have been proposed to test for quintessence cosmologies (e.g. \citealt{lind}) since the equation of state of the quintessence component and its evolution influence the value of the Hubble constant. Although the differences between evolving and non-evolving cosmologies are relatively small, and an order of magnitude more lenses than currently available would be needed to accurately probe any quintessence component 
 \citep{lewi}, this will be possible in the near future.

We have seen how lensing time delays have the potential to constrain the $\Omega_{\rm m}$ -- $\Omega_{\Lambda}$ plane.  However, this analysis depended in part upon the assumption of Gaussian distributions for the bulk properties of sources and lenses in future lens survey samples.  For some of the parameters this only served as a necessary approximation.  Indeed, the nature of the distribution of density slopes, for example, could well have pronounced tails in the ``super-isothermal'' region.  While it is true that the average profile of the ensemble may be isothermal, a number of systems exhibiting super-isothermality could result in non-Gaussian distributions.  Similarly, features such as ellipticity distributions are not particularly well defined for lens galaxies at present.  Future work on trying to extract cosmological information from large lens time delay samples would do well to explore the effects of more realistic, non-Gaussian distributions.
 
In this study we have demonstrated what can be expected in the way of constraints through the use of a commonly adopted elliptical model approach.  Different methods such as the non-parametric approach to modelling lens systems (e.g. \citealt{saha}), or indeed statistical methods involved in $H_{0}$ derivations (e.g. \citealt{ogur}) could lead to greater accuracy in the constraints of parameters studied here, or similar constraints with smaller sample sizes.  Whichever method is used in the approach, the new data samples from future surveys promise to yield important results from strong lensing systems and as such are eagerly anticipated.

\section*{Acknowledgments}

Thanks to Richard Ellis, Leonidas Moustakas and Dan Coe for discussions relating to this work.  This work was supported by PPARC through a PhD studentship (BMD) and by the Royal Society (LJK).   This work was supported in part by the European Community's Sixth Framework Marie Curie Research Training Network Programme, Contract No. MRTN-CT-2004-505183 ``ANGLES''.

\end{document}